% =======================LETRAS HUECAS============================
\newfam\msbfam
\font\twlmsb=msbm10 at 12pt
\font\eightmsb=msbm10 at 8pt
\font\sixmsb=msbm10 at 6pt
\textfont\msbfam=\twlmsb
\scriptfont\msbfam=\eightmsb
\scriptscriptfont\msbfam=\sixmsb
\def\cj{\fam\msbfam}

\def\C{{\cj C}}

\def\R{{\cj R}}

\centerline{\bf EPR, Bell, GHZ, and Hardy theorems, and quantum mechanics $^{a)}$}

\

\centerline{\bf M. Socolovsky} 

\

\centerline{\it Instituto de Ciencias Nucleares, Universidad Nacional Aut\'onoma de M\'exico}
\centerline{\it Circuito Exterior, Ciudad Universitaria, 04510, M\'exico D. F., M\'exico} 

\

{\it We review the theorems of Einstein-Podolsky-Rosen (EPR), Bell, Greenberger-Horne-Zeilinger (GHZ), and Hardy, and present arguments supporting the idea  that quantum mechanics is a complete, causal, non local, and non separable theory.}

\

{\bf 1. Introduction}

\

In 1935, Einstein, Podolsky and Rosen (EPR) published the today 70 years old paper ``Can Quantum-Mechanical Description of Physical Reality Be Considered Complete$^1$, in which they gave a negative answer to this question. The incompleteness refers to an individual quantum system, not to an ensemble of identical systems (section 2). To arrive at this conclusion, EPR assumed locality and a criterion for the result of a measurement of a physical quantity to be considered an element of physical reality prior to the measurement, and established a necessary condition for a physical theory to be considered complete. This led to the search of a ``complete theory'' by adding ``hidden'' variables to the wave function in order to implement realism, the most celebrated of this kind of theories being the De Broglie-Bohm theory.$^{27}$

It was not until 29 years later, that Bell published his famous paper ``On the Einstein-Podolsky-Rosen paradox''$^{13}$ in which, taking as starting points the EPR hypotesis plus the assumption of the existence of hidden variables, he derived an inequality between two-particle correlation functions (averages of products of spin projections along arbitrary directions in the case of atoms, nucleons and electrons, or polarizations in the case of photons) which for some range of angles or polarizations is violated by the quantum formulae for the correlations (section 3). For a particular value of the angle between the spin directions or of photon polarizations, as in the EPR case, the correlation is perfect, that is, the result of the measurement on one particle allows the prediction with certainty -{\it i.e.} with probability equal to 1- of the result of the measurement on the other particle, coincides with the EPR result and does not contradict the quantum prediction; for the other values of angles or polarizations the correlation is imperfect {\it i.e.} only probabilities -less than 1- of the outcomes of the measurements on the 2nd. particle are predicted from the result of the measurement on the 1st. particle. The Bell's result offered the possibility to decide experimentally -and theoretically if one believes a priori in the quantum formulae-between the orthodox view$^5$ of quantum mechanics and the interpretation based on the EPR hypotesis plus, possibly, hidden variables. The experimental evidence in the last decades$^{15-18}$ favours the first alternative. 

Within the framework of quantum mechanics one can give arguments to support the statement that the theory is {\it non separable} (subsection 2.1.a) and {\it relativistically causal} but {\it non local} in the sense of EPR-Bell (subsection 2.2.g). Then an analysis of the Greenberger-Horne-Zeilinger theorem (``Bell's theorem without inequalities'') of 1989 shows a contradiction between quantum mechanics and local realism, but for perfect correlations of three or more particles. One concludes that quantum mechanics is a {\it complete theory} in the sense that it provides the most complete description of a quantum system, without appealing to the concept of elements of reality and/or to hidden variables (section 4). 

Hardy's theorem (1992) (section 5) is a Bell's theorem for perfect correlations between two particles {\it i.e.} establishes a contradiction between quantum mechanics and local realism for this system, and shows that EPR elements of reality corresponding to Lorentz invariant observables are not Lorentz invariant. However, since quantum mechanics is causal, no signal can be sent backwards in time and therefore there is no need of a special reference frame to block causal paradoxes. 

\

{\bf 2. EPR theorem (1935)}

\

The EPR theorem$^1$ says that {\it quantum mechanics does not provide a complete description of an individual quantum system}. This result, commonly known as the EPR {\it paradox} since Schroedinger$^2$ considered as such the situation already pointed out in the original paper by EPR, namely that to a distant physical system one could associate two (or more) different quantum states or wave functions, was derived assuming the validity of the following hypotesis:

i) Quantum mechanics (QM) provides a correct description of nature.

ii) Locality: ``..., since at the time of measurement the two systems no longer interact, no real change can take place in the second system in consequence of anything that may be done to the first system.''$^1$ or ``But on one suposition we should, in my opinion, absolutely hold fast: the real factual situation of the system $S_2$ is independent of what is done with the system $S_1$, which is spatially separated from the former.''$^3$

iii) Physical reality (sufficient condition): ``If, without in any way disturbing a system, we can predict with certainty (i.e. with probability equal to unity) the value of a physical quantity, then there exists an element of physical reality corresponding to this physical quantity''$^1$. This concept, from the physical point of view, is rather obscure, since ``physical reality'' was not defined before; it rests rather on intuition.

iv) Completeness of a theory: any element of the physical reality has to have a counterpart in the theory. 

In the literature, the set of hypotesis ii) and iii) is called {\it local realism}. 

\

{\it Proof of the theorem}

\

We shall consider here the finite dimensional Hilbert space approach of Bohm$^4$, who replaced the continuum spectrum variables $p$ and $q$ by the spin ${{1}\over {2}}$ components of a pair of particles. A particle of total spin $S=0$ decays into two spin ${{1}\over{2}}$ particles of the same mass, which move appart in opposite directions; according to quantum mechanics, the spin wave function of the pair of particles is given by $$\psi={{1}\over{\sqrt{2}}}(\uparrow_1 \otimes \downarrow_2-\downarrow_1 \otimes \uparrow_2) \eqno{(2.1)}$$ where $\uparrow_k$ ($\downarrow_l$) is the wave function of particle $k$ ($l$) with spin projection $+{{1}\over{2}}$ (-${{1}\over{2}}$) in the space direction specified by the unit vector $\hat{n}=(sin \theta cos \varphi, sin \theta sin \varphi, cos \theta)$. A more precise notation is $$\vert \psi>={{1}\over{\sqrt{2}}}(\vert \hat{n}>_1 \otimes \vert-\hat{n}>_2- \vert -\hat{n}>_1\otimes \vert \hat{n}>_2) \eqno{(2.1')}$$ where $\vert \hat{n}>_l$ are kets in the Hilbert space of each particle. In terms of the polar angles $\theta$ and $\varphi$, $$\vert \hat{n}>=cos{{\theta}\over {2}}\vert \uparrow>+e^{i\varphi}sin{{\theta}\over{2}}\vert \downarrow>, \ \ \ \vert -\hat{n}>=sin{{\theta}\over{2}}\vert \uparrow>-e^{i\varphi}cos{{\theta}\over{2}}\vert \downarrow>, \eqno{(2.1^{''})}$$ where $\vert \uparrow>$ and $\vert \downarrow>$ are spin states in directions $\hat{z}$ and $-\hat{z}$ respectively. Notice however that this does not mean that each particle is in a pure state (see below 2.2.f.). A state like (2.1) which is not a product of states but a sum of products of states, is called an {\it entangled} state. According to Schroedinger$^2$, entanglement is ``the characteristic trait of quantum mechanics''. {\it If one measures} the spin component of particle 1 with a Stern-Gerlach (SG) apparatus in the direction $\hat{m}$ and finds $+{{1}\over{2}}$ ($-{{1}\over{2}}$), then one can predict with certainty that the spin component in the same direction of particle 2 is $-{{1}\over{2}}$ ($+{{1}\over {2}}$). Since by assumption the two particles do not interact, according to the hypotesis ii) and iii) the spin component of particle 2 is an element of physical reality. It is important to realize that this way of thinking implies that $\psi$ has suffered one of the following two {\it collapses}: $$\psi \to \uparrow_1 \otimes \downarrow_2 \ \ \ or \ \ \ \psi \to -\downarrow_1 \otimes \uparrow_2. \eqno{(2.2)}$$ But the choice of direction $\hat{m}$ in the measurement of particle 1 was arbitrary, and one {\it could} have chosen for example the directions $\hat{z}$, $\hat{y}$ or $\hat{x}$, respectively giving physical reality to the spin components $S_z$, $S_y$ or $S_x$ of particle 2. But according to QM, since the spin operators $\hat{S}_k$, $k=x,y,z$ do not commute and satisfy $$[\hat{S}_k,\hat{S}_l]=i\hat{S}_m \eqno{(2.3)}$$ with $k,l,m$ in cyclic order, only one component can have a precise value. Then QM does not provide a complete description of the physical reality of any of the two particles.    QED   

\

2.1. {\it Some definitions}

\

2.1.a. A quantum system consisting of $N$ non interacting parts is called {\it separable} if its wave function $\psi$ is the product of the wave functions $\psi_i$ of its constituent parts $i=1,...,N$: $$\psi=\psi_1...\psi_N,$$ independently whether these parts have interacted or not among themselves before.

The existence of entangled states like (2.1) shows that {\it QM is a non separable theory}. 

It is clear that a separable theory is local in the sense of ii) above, but not the other way around, that is, separability is a stronger property than locality. Then non separability does not imply non locality: a non separable theory can be local or non local. 

\

2.1.b. By {\it relativistic causality} or simply {\it causality}, we understand that no signal (or information) can travel faster than light in the vacuum. This is equivalent to the statement that a particle can be acted only by other particles or fields lying in its past light cone. A theory can be non local (and then non separable) but however causal. This is the case of QM (see 2.2.g.).

\

2.2. {\it Observations and criticisms}

\

2.2.a. The EPR theorem {\it does not contradict} neither QM (only claims its incompleteness) nor mathematics.

\

2.2.b. The EPR {\it correlation} between particles is {\it perfect}, that is, the conditional probability of the result of the measurement on the second particle, given the probability (${{1}\over {2}}$) that the result of the measurement on the first particle is $+{{1}\over{2}}$ or $-{{1}\over{2}}$, is 1 {\it i.e. certainty}: this is a result of QM, and in this lenguage, related to the concept of collapse, one can write (with obvious notation) $$p(\uparrow_1,\downarrow_2)=p(\downarrow_2\vert \uparrow_1)\times p(\uparrow_1)=1\times {{1}\over {2}}={{1}\over{2}} \eqno{(2.4)}$$ where the $p's$ denote probabilities, and $p(A\vert B)$ denotes the conditional probability of $A$ given $B$.

\

2.2.c. A direct calculation in QM shows that, without passing through the intermediate steps of first computing a probability and then a conditional probability, in the state (2.1) the {\it joint probability} for, say $\vert+\hat{m}>_1\otimes \vert-\hat{m}>_2$ is given by $$p(\vert\hat{m}>_1,\vert-\hat{m}>_2)={{1}\over {2}}. \eqno{(2.5)}$$ In fact, the probability amplitude for finding, in the singlet state (2.1), particle 1 (2) with spin projection in direction $\hat{a}$ ($\hat{b}$) is $$A(\vert\hat{a}>_1, \vert\hat{b}>_2)=
{{1}\over{\sqrt{2}}} \ _1<\hat{a}\vert \otimes \ _2<\hat{b}\vert(\vert \hat{n}>_1\otimes\vert -\hat{n}>_2-\vert -\hat{n}>_1\otimes\vert\hat{n}>_2)$$ $$={{1}\over{\sqrt{2}}}(_1<\hat{a}\vert\hat{n}>_1 \ _2<\hat{b}\vert -\hat{n}>_2- \ _1<\hat{a}\vert -\hat{n}>_1 \ _2<\hat{b}\vert \hat{n}>_2)$$ $$={{1}\over{\sqrt{2}}}((cos{{\theta_a}\over{2}}<\uparrow\vert +e^{-i\varphi_a}sin{{\theta_a}\over{2}}<\downarrow \vert)(cos{{\theta_n}\over{2}}\vert \uparrow>+e^{i\varphi_n}sin{{\theta_n}\over{2}}\vert\downarrow>)$$ $$\times(cos{{\theta_b}\over{2}}<\uparrow\vert +e^{-i\varphi_b}sin{{\theta_b}\over{2}}<\downarrow\vert )(sin{{\theta_n}\over {2}}\vert \uparrow>-e^{i\varphi_n}cos{{\theta_n}\over{2}}\vert \downarrow >)$$ $$-(cos{{\theta_a}\over {2}}<\uparrow\vert +e^{-i\varphi_a}sin{{\theta_a}\over{2}}<\downarrow\vert)(sin{{\theta_n}\over{2}}\vert\uparrow>-e^{i\varphi_n}cos{{\theta_n}\over{2}}\vert\downarrow>)$$ $$\times (cos{{\theta_b}\over{2}}<\uparrow\vert +e^{-i\varphi_b}sin{{\theta_b}\over{2}}<\downarrow\vert)(cos{{\theta_n}\over{2}}\vert\uparrow>+e^{i\varphi_n}sin{{\theta_n}\over{2}}\vert\downarrow>))$$ $$={{1}\over{\sqrt{2}}}((cos{{\theta_a}\over{2}}sin{{\theta_n}\over{2}}+e^{-i(\varphi_a-\varphi_n)}sin{{\theta_a}\over{2}}sin{{\theta_n}\over{2}})(cos{{\theta_b}\over{2}}sin{{\theta_n}\over{2}}-e^{-i(\varphi_b-\varphi_n)}sin{{\theta_b}\over{2}}cos{{\theta_n}\over{2}})$$  $$-(cos{{\theta_a}\over{2}}sin{{\theta_n}\over{2}}-e^{-(\varphi_a-\varphi_n)}sin{{\theta_a}\over{2}}cos{{\theta_n}\over{2}})(cos{{\theta_b}\over{2}}cos{{\theta_n}\over{2}}+e^{-i(\varphi_b-\varphi_n)}sin{{\theta_b}\over{2}}sin{{\theta_n}\over{2}}))$$  $$={{1}\over{\sqrt{2}}}(e^{-i(\varphi_a-\varphi_n)}sin{{\theta_a}\over{2}}cos{{\theta_b}\over{2}}-e^{-i(\varphi_b-\varphi_n)}sin{{\theta_b}\over{2}}cos{{\theta_a}\over{2}}), \eqno{(2.6)}$$ which gives the probability $$p(\vert\hat{a}>_1,\vert\hat{b}>_2)=\vert A(\vert\hat{a}>_1,\vert \hat{b}>_2) \vert ^2={{1}\over{2}}(sin^2{{\theta_a}\over{2}}cos^2{{\theta_b}\over{2}}+sin^2{{\theta_b}\over {2}}cos^2{{\theta_a}\over{2}}-{{1}\over {2}}sin \theta_a sin \theta_b cos(\varphi_a-\varphi_b)). \eqno{(2.7)}$$ For the particular case $\hat{a}=\hat{m}$ and $\hat{b}=-\hat{m}$, (2.7) leads to (2.5). For the other cases $p(\vert -\hat{m}>_1,\vert\hat{m}>_2)={{1}\over{2}}$, $p(\vert \hat{m}>_1,\vert \hat{m}>_2)=p(\vert -\hat{m}>_1,\vert -\hat{m}>_2)=0$.) This method does {\it not} require the concept of collapse and is at the basis of the {\it orthodox position} with respect to QM: (2.5) or, in general, (2.6) and (2.7), is the {\it unique} information that QM provides on the physical system and so the conclusion of incompleteness of QM by EPR and the idea of an action at a distance involving superluminal transmission of signals are {\it false}. 

\

2.2.d. Moreover, one of the most powerful criticisms to the arguments of EPR is the meaning of the words ``can predict'' in the hypotesis iii). As employed by EPR, these words are used in the {\it weak sense}, which means that it is enough with the possibility of making the measurements, even if they are not performed, that the realities of the physical quantities in the remote system occur ({\it counterfactual} alternative). The {\it strong sense} however, requires that the measurements on the first system have to be actually performed and this is not possible if the observables in question have no common eigenvalue, as for example $S_x^{(1)}$ and $S_z^{(1)}$. In this case, $S_x^{(2)}$ and $S_z^{(2)}$ can not be elements of the physical reality and the conclusion of incompleteness is invalidated. This is essentially the criticism of Bohr$^5$; see also the references 6 and 7. {\it This argument allows the coexistence of QM with relativistic causality, but destroys the EPR sufficient criterion of reality.}

\

2.2.e. We shall see that the EPR hypotesis ii) and iii), for statistical correlations (Bell's theorem) between two particles (section 3), and for perfect correlations between three (or more) particles (section 4) and for two particles (section 5) contradict QM.

\

2.2.f. The most general description of the state of a quantum mechanical system is through the {\it density or statistical operator} $\hat{\rho}$. If the system is in a {\it pure state}, then its description is given by a normalized ket $\vert \psi> \in {\cal H}$ (${\cal H}$ is the Hilbert space), and $\hat{\rho}$ is given by $$\hat{\rho}=\vert\psi>\otimes <\psi\vert \in {\cal H}\otimes {\cal H}^*, \eqno{(2.8)}$$ ${\cal H}^*$ being the dual space of ${\cal H}$. Clearly, $$\hat{\rho}^2=\hat{\rho}. \eqno{(2.8.a)}$$

Instead, a {\it mixed state} is described by the operator $$\hat{\rho}=\Sigma_i \ w_i \vert\psi_i>\otimes <\psi_i \vert \eqno{(2.9)}$$ with $w_i \in (0,1)$ and $\Sigma_iw_i=1$ (the different normalized kets $\vert\psi_i>$ need not be orthogonal). If $\{\vert a>\}$ is a complete set, then $$tr(\hat{\rho}) =\Sigma_a \ <a\vert \hat{\rho}\vert a>=\Sigma_a\Sigma_i \ w_i<a\vert\psi_i><\psi_i\vert a>= \Sigma_i \ w_i\Sigma_a\vert<\psi_i\vert a>\vert^2=\Sigma_i \ w_i\vert \vert \psi_i \vert \vert^2=\Sigma_i w_i=1. \eqno{(2.10)}$$ Then $$tr(\hat{\rho}^2)=1 \ \ \ for \ a \ pure \ state. \eqno{(2.11)}$$ Clearly, $\hat{\rho}^{\dagger}=\hat{\rho}$ {\it i.e.} $\hat{\rho}$ is {\it hermitian}, and $<\psi\vert \hat{\rho}\vert \psi> \geq 0$ for all $\vert \psi>\in {\cal H}$ {\it i.e.} $\hat{\rho}$ is a {\it non negative} operator with eigenvalues satisfying $1\geq \rho_k \geq 0$ and $\Sigma_k \rho_k=1$. Then $$tr(\hat{\rho}^2)<1 \ \ \ if \ and \ only \ if \ \hat{\rho} \ represents \ a \ mixed \ state. \eqno{(2.12)}$$ Both for a pure or mixed state, the average value of an observable $A$ is given by $$<\hat{A}>=tr(\hat{\rho}\hat{A}) \eqno{(2.13)}$$ where $\hat{A}$ is the hermitian operator correponding to $A$. In fact, if $\{\vert a>\}$ are its eigenkets, then $$<\hat{A}>=\Sigma_i \ w_i <\hat{A}>_i=\Sigma_i \ w_i <\psi_i\vert \hat{A} \vert \psi_i>=\Sigma_i\Sigma_a \Sigma_{a^\prime} \ w_i<\psi_i\vert a^\prime><a\vert \psi_i><a^\prime\vert \hat{A}\vert a>$$ $$=\Sigma_i \Sigma_a \Sigma_{a^\prime} \ w_i <\psi_i\vert a^{\prime}><a\vert\psi_i>a<a^\prime\vert a>=\Sigma_i\Sigma_a \Sigma_{a^\prime} \ w_i <\psi_i\vert a^\prime><a\vert \psi_i>a \delta_{aa^\prime}$$ $$= \Sigma_i\Sigma_a \ w_i<\psi_i\vert a><a\vert \psi_i>a=\Sigma_a<a\vert (\Sigma_i \ w_i\vert \psi_i>\otimes <\psi_i\vert )\hat{A}\vert a>=\Sigma_a<a\vert\hat{\rho}\hat{A}\vert a>.$$ Also, $$<a\vert \hat{\rho}\vert a>=<a\vert (\Sigma_i \ w_i\vert \psi_i>\otimes <\psi_i\vert)\vert a>=\Sigma_i \ w_i\vert<a\vert \psi_i>\vert ^2=p(a). \eqno{(2.14)}$$ Finally, it can be easily shown that since each $\vert \psi_i>$ satisfies the Schroedinger equation, then $\hat{\rho}$ satisfies $${{d\hat{\rho}}\over {dt}}=i[\hat{\rho},\hat{H}] \eqno{(2.15)}$$ where $\hat{H}$ is the hamiltonian. So $\hat{\rho}$ evolves with time with an opposite sign as the evolution of the Heisenberg picture of a time independent (in the Schroedinger picture) observable: ${{d\hat{A}_H}\over {dt}}=-i[\hat{A}_H, \hat{H}]$.
 
\

Consider now a quantum system consisting in two parts, 1 and 2. Let $\theta_1$ be an observable for part 1; then its average value in the state of the total system described by the density operator $\hat{\rho}_{12}$ is given by $$<\hat{\theta}_1\otimes I_2>_{12}=tr_{12}(\hat{\theta}_1\otimes I_2 \hat{\rho}_{12})=tr_1 tr_2(\hat{\theta}_1 \otimes I_2 \hat{\rho}_{12})=tr_1(\hat{\theta}_1tr_2 (I_2\hat{\rho}_{12}))=tr_1(\hat{\theta}_1 \hat{\rho}_1^{red}) \eqno{(2.16)}$$ where $$\hat{\rho}_1^{red}=tr_2(\hat{\rho}_{12}) \eqno{(2.16a)}$$ is the {\it reduced density operator} for the subsystem 1; similarly $$<I_1\otimes \hat{\theta}_2>=tr_2(\hat{\theta}_2 \hat{\rho}_2^{red}) \eqno{(2.17)}$$ with $$\hat{\rho}_2^{red}=tr_1(\hat{\rho}_{12}). \eqno{(2.17a)}$$ ($I_k, \ k=1,2$ are the identity operators on the corresponding Hilbert spaces.) In particular, for the pure state (2.1), $$\hat{\rho}^{red}_1=<\uparrow_2\vert \psi>\otimes <\psi\vert \uparrow_2>+<\downarrow_2\vert \psi>\otimes<\psi\vert \downarrow_2>$$ $$={{1}\over{2}}<\uparrow_2\vert(\vert\uparrow_1>\otimes\vert\downarrow_2>-\vert\downarrow_1>\otimes\vert\uparrow_2>)\otimes(<\uparrow_1\vert\otimes<\downarrow_2\vert -<\downarrow_1\vert\otimes<\uparrow_2\vert)\vert\uparrow_2>$$ $$+{{1}\over{2}}<\downarrow_2\vert (\vert\uparrow_1>\otimes\vert\downarrow_2>-\vert\downarrow_1>\otimes\vert\uparrow_2>)\otimes (<\uparrow_1\vert\otimes <\downarrow_2\vert-<\downarrow_1\vert\otimes<\uparrow_2\vert )\vert\downarrow_2>$$ $$={{1}\over{2}}(\vert\uparrow_1>\otimes<\uparrow_1\vert+\vert\downarrow_1>\otimes<\downarrow_1\vert ), \eqno{(2.18a)}$$ and similarly, $$\hat{\rho}^{red}_2={{1}\over{2}}(\vert\uparrow_2>\otimes<\uparrow_2\vert +\vert\downarrow_2>\otimes<\downarrow_2\vert). \eqno{(2.18b)}$$ Therefore, in a matrix representation, $$\rho^{red}_1=\rho^{red}_2={{1}\over{2}}\pmatrix{1 & 0 \cr 0 & 1 \cr}={{1}\over{2}}I. \eqno{(2.19)}$$ So, $$(\rho^{red}_k)^2={{1}\over{4}}I\neq {{1}\over{2}}I=\rho^{red}_k, \ k=1,2$$ and $$tr((\rho^{red}_k)^2)={{1}\over{2}}<1 \eqno{(2.20)}$$ {\it i.e.} each of the entangled spins in the pure state $\psi$ is {\it not} in a pure state, but in a mixed state: the whole system is described by a wave function, but not each of its parts.

An immediate consequence of this fact is that the average value of any spin component of each particle vanishes: $$<\sigma_k>_l=tr(\sigma_k\rho_l^{red})=0, \ k=1,2,3, \ l=1,2, \eqno{(2.21)}$$ ($\sigma_k$ are the Pauli matrices $\sigma_1=\pmatrix{0 & 1 \cr 1 & 0 \cr}$, $\sigma_2=\pmatrix{0 & -i \cr i & 0 \cr}$ and $\sigma_3=\pmatrix{1 & 0 \cr 0 & -1 \cr}$), which means that no spin component of any of the two particles has a definite or precise value (and therefore a reality) that can be predicted with certainty without perturbing the corresponding particle (subsystem)$^{8}$.   

\

2.2.g. {\it Impossibility of an instantaneous (and therefore superluminal) sending of information. Non locality.}

In the situation described by the state (2.1), assume that observer 1 finds $\uparrow$, then observer 2 finds $\downarrow$. If this would imply a transmision of information from 1 to 2 and not only a correlation between 1 and 2, observer 2 should be able to distinguish between the following alternatives:

i) 1 found $\uparrow$,

ii) 1 did not measure,

iii) 1 measured in an arbitrary direction $\hat{n}$.

It is clear that 2 can not determine what actually ocurred at 1, but only if the two observers come together and compare their results. However, though 1 can {\it not} send an information to 2 and therefore {\it causality} (2.1.b.) {\it is not violated}, as a consequence of the {\it individual} measurement at 1, {\it the probability distribution at 2 has changed instantaneously} since, from (2.14), $$p(\uparrow_1)=<\uparrow_1\vert\hat{\rho}^{red}_1\vert\uparrow_1>=<\uparrow_1\vert {{1}\over{2}}I\vert\uparrow_1>={{1}\over{2}}<\uparrow_1\vert\uparrow_1>={{1}\over{2}},$$ and from (2.5), $$p(\downarrow_2\vert\uparrow_1)={{p(\uparrow_1,\downarrow_2)}\over{p(\uparrow_1)}}={{{1}\over{2}}\over{{1}\over{2}}}=1 \ (certainty). \eqno{(2.22)}$$ If 1 would not have measured, then $$p(\downarrow_2)=<\downarrow_2\vert\hat{\rho}^{red}_2\vert\downarrow_2>=<\downarrow_2\vert{{1}\over{2}}I\vert\downarrow_2>={{1}\over{2}}<\downarrow_2\vert\downarrow_2>={{1}\over{2}}.$$ This amounts to an {\it instantaneous change of state} of 2 ({\it collapse}), from a mixed state to a pure state {\it i.e.} to an instantaneous (and therefore superluminal) change of the density operator: $$\hat{\rho}^{red}_2 \to \vert \downarrow_2>$$ and therefore to a violation of locality {\it a` la} \ EPR. Since this is a pure quantum mechanical result, we conclude that {\it QM is a non local theory.}

In the following we shall see that, {\it in the average}, the above instantaneous change of probability distribution does not occur.

\

2.2.h. {\it Impossibility, in the average, of an instantaneous (and therefore superluminal) modification of a remote probability distribution}$^{9,10}$

Let $U$ and $V$ denote the two non interacting and spatially separated parts of a composite quantum system ``$U+V$'' (we can assume that in the past the two parts interacted with each other), and $A$ and $B$ corresponding measuring apparatuses; the total Hilbert is given by $${\cal H}={\cal H}_A\otimes {\cal H}_U\otimes{\cal H}_V\otimes{\cal H}_B. \eqno{(2.23)}$$ The interaction hamiltonians apparatuses-subsystems, respectively $H(U,A)$ and $H(V,B)$, lead to unitary evolution operators ${\cal U}(U,A)\equiv {\cal U}_{UA}$ and ${\cal U}(V,B)\equiv {\cal U}_{VB}$ which by causality commute with each other: $$[{\cal U}_{UA},{\cal U}_{VB}]=0. \eqno{(2.24)}$$ ${\cal U}_{UA}$ (${\cal U}_{VB}$) is the identity on ${\cal H}_{VB}={\cal H}_V\otimes {\cal H}_B$ (${\cal H}_{UA}={\cal H}_U\otimes {\cal H}_A$). Before any measurement, the density operator of the total system is $$\hat{\rho}^0=\hat{\rho}_A^0\otimes \hat{\rho}^0_{UV}\otimes\hat{\rho}^0_B, \eqno{(2.25)}$$ and in general $\hat{\rho}^0_{UV}\neq\hat{\rho}^0_U\otimes\hat{\rho}^0_V$. If $A$ and $B$ perform succesive measurements, the density operator evolves to $$\hat{\rho}_{BA}=({\cal U}_{VB}{\cal U}_{UA})\hat{\rho}^0({\cal U}_{VB}{\cal U}_{UA})^\dagger={\cal U}_{VB}{\cal U}_{UA}\hat{\rho}^0_A\otimes \hat{\rho}^0_{UV}\otimes\hat{\rho}^0_B{\cal U}^\dagger_{UA}{\cal U}^\dagger_{VB}. \eqno{(2.26)}$$ On the other hand, if the apparatus $A$ is switched off, {\it i.e.} $A$ makes no measurement, then ${\cal U}_{UA}$ is the identity (since $H(U,A)=0$): ${\cal U}_{UA}=I_U\otimes I_A$. 

Let $\theta_V$ be an observable of the subsystem $V$; for its expectation value one has $$<\hat{\theta}_V>_{\hat{\rho}_{BA}}=tr(\hat{\theta}_V\hat{\rho}_{BA})=tr(\hat{\theta}_V{\cal U}_{VB}{\cal U}_{UA} \hat{\rho}^0_A\otimes\hat{\rho}^0_{UV}\otimes\hat{\rho}^0_B{\cal U}^\dagger_{UA}{\cal U}^\dagger_{VB})$$ $$=tr({\cal U}^\dagger_{UA}{\cal U}^\dagger_{VB}\hat{\theta}_V{\cal U}_{VB}{\cal U}_{UA}\hat{\rho}^0_A\otimes \hat{\rho}^0_{UV}\otimes \hat{\rho}^0_B);$$ since by locality, $[\hat{\theta}_V, {\cal U}_{UA}]=0$, one has $$<\hat{\theta}_V>_{\hat{\rho}_{BA}}=tr({\cal U}^\dagger_{VB}\hat{\theta}_V{\cal U}_{VB}\hat{\rho}^0_A\otimes \hat{\rho}^0_{UV}\otimes \hat{\rho}^0_B)=tr(\hat{\theta}_V({\cal U}_{VB}\hat{\rho}^0_A\otimes \hat{\rho}^0_{UV}\otimes \hat{\rho}^0_B {\cal U}^\dagger_{VB}))= tr(\hat{\theta}_V\hat{\rho}_B)=<\hat{\theta}_V>_{\hat{\rho}_B}. \eqno{(2.27)}$$ That is, the mean value of the observable $\theta_V$ of the subsystem $V$ does not depend whether a measurement with the apparatus $A$ is performed or not on the subsystem $U$. 

Consider in particular $\hat{\theta}_V=\vert\beta>\otimes<\beta \vert$, the projector associated with the eigenvlue $\beta$ of an observable $\hat{\beta}$ of the subsystem $V$; from (2.13) its mean value is given by $$tr((\vert \beta>\otimes<\beta\vert )\hat{\rho}_{BA})=tr((\vert \beta>\otimes <\beta\vert )\hat{\rho}_B)$$ and therefore $$\Sigma_{\beta^\prime} \ <\beta^\prime\vert\beta><\beta\vert\hat{\rho}_{BA}\vert \beta^\prime>=\Sigma_{\beta^\prime}<\beta^\prime\vert\beta><\beta\vert \hat{\rho}_B\vert \beta^\prime>$$ {\it i.e.} $$<\beta\vert\hat{\rho}_{BA}\vert\beta>=<\beta\vert\hat{\rho}_B\vert\beta>. \eqno{(2.28)}$$ Then, from (2.14), $$p_{BA}(\beta)=p_B(\beta). \eqno{(2.29)}$$ So, in contradistinction with the result for a measurement performed on an individual quantum system (2.2.g.), (2.28) expresses the fact that {\it at the statistical level}, that is, {\it at the ensamble level}, the probability distributions remain unaltered by measurements on a distant subsystem.

\

2.2.i. Formulae analogous to (2.6) and (2.7), but for photons, can be easily derived$^{11,12}$; these formulae are crucial for the comparison of the predictions of QM and the EPR and Bell theorems. 

Let $\vert H_l>$ and $\vert V_l>$ represent respectively the states of horizontal and vertical photon polarizations in an EPR type experiment ($l=1,2$), and $$\pmatrix{\vert\theta_l> \cr \vert\theta_l^\perp >\cr}=\pmatrix{cos\theta_l & sin\theta_l \cr -sin\theta_l & cos\theta_l \cr}\pmatrix{\vert H_l> \cr \vert V_l> \cr} \eqno{(2.30)}$$ with $\theta_l^\perp=\theta_l+{{\pi}\over{2}}$, the polarization states corresponding to the ordinary ($\theta_l$) and extraordinary ($\theta_l^\perp$) photon ``paths'' in a calcite crystal detector. Then, the two-photon states $$\vert\psi_I>={{1}\over \sqrt{2}}(\vert V_1>\otimes\vert H_2>-\vert H_1>\otimes\vert V_2>), \eqno{(2.31)}$$ decay product of the ground state of positronium (negative parity), and $$\vert\psi_{II}>={{1}\over\sqrt{2}}(\vert H_1>\otimes\vert H_2>+\vert V_1>\otimes\vert V_2>), \eqno{(2.32)}$$ coming from a cascade process $J=0\to 1\to 0$ in calcium atoms (positive parity), with the help of (2.30) can be written as follows: $$\vert\psi_I>=\Sigma_{a,b\in\{1,\perp\}} \ A_{12}^{ab}(I)\vert\theta_1^a>\otimes\vert\theta_2^b>, \eqno{(2.31')}$$ $$\vert\psi_{II}>=\Sigma_{a,b\in\{1,\perp\}} \ A_{12}^{ab}(II)\vert\theta_1^a>\otimes\vert\theta_2^b>, \eqno{(2.32')}$$ where $\theta_l^1\equiv\theta_l$, $l=1,2$, and $$A_{12}^{11}(I)=A_{12}^{\perp\perp}={{1}\over\sqrt{2}}sin(\theta_1-\theta_2), \eqno{(2.33a)}$$ $$A_{12}^{\perp 1}(I)=-A_{12}^{1\perp}(I)={{1}\over\sqrt{2}}cos(\theta_1-\theta_2), \eqno{(2.33b)}$$ $$A_{12}^{11}(II)=A_{12}^{\perp\perp}(II)={{1}\over\sqrt{2}}cos(\theta_1-\theta_2), \eqno{(2.34a)}$$ $$A_{12}^{1\perp}(II)=-A_{12}^{\perp 1}(II)=-{{1}\over\sqrt{2}}sin(\theta_1-\theta_2). \eqno{(2.34b)}$$ Then, the corresponding {\it joint probabilities}, again the {\it unique} prediction of QM, are: $$p_I(\vert\theta_1>,\vert\theta_2>)=p_I(\vert\theta_1^\perp>,\vert\theta_2^\perp>)=p_{II}(\vert\theta_1^\perp>,\vert\theta_2>)=p_{II}(\vert\theta_1>,\vert\theta_2^\perp>)={{1}\over{2}}sin^2(\theta_1-\theta_2), \eqno{(2.35a)}$$ and $$p_I(\vert\theta_1^\perp>,\vert\theta_2>)=p_I(\vert\theta_1>,\vert\theta_2^\perp>)=p_{II}(\vert\theta_1>,\vert\theta_2>)=p_{II}(\vert\theta_1^\perp>,\vert\theta_2^\perp>)={{1}\over{2}}cos^2(\theta_1-\theta_2).\eqno{(2.35b)}$$ Clearly, $$\Sigma_{a,b\in\{1,\perp\}}\vert \ A_{12}^{ab}(I)\vert^2=\Sigma_{a,b\in\{1,\perp\}}\vert A_{12}^{a,b}(II)\vert^2=1 \eqno{(2.36)}$$ {\it i.e.} $$\Sigma_{a,b\in\{1,\perp\}} \ p_I(\vert\theta_1^a>,\vert\theta_2^b>)=\Sigma_{a,b\in\{1,\perp\}} \ p_{II}(\vert\theta_1^a>, \vert\theta_2^b>)=1. \eqno{(2.36')}$$

\

{\bf 3. Bell's theorem$^{13}$ (1964)}

\

The simplest statement of the theorem is the following: {\it Quantum mechanics violates local realism}. The incompleteness of QM as claimed by EPR for the description of an individual quantum system, led to the idea that the incorporation of {\it additional variables} $\lambda$, called {\it hidden variables}, could complete the theory. They can be arbitrary in number (in general finite) and should explain realism. 

\

{\it Proof of Bell's theorem}

\

The hypotesis are the same as for the EPR theorem, {\it plus} the assumption of the existence of hidden variables $\lambda$, such that the pair $$(\psi,\lambda)$$ gives a complete description of the system. As in EPR, take the system described by the wave function (2.1) and let the SG which measures the projection of spin 1 (2) be in the direction specified by the unit vector $\hat{a}$ ($\hat{b}$) of $\R^3$. One assumes that in the laboratory both measurements are simultaneous and that the choice of $\hat{a}$ and $\hat{b}$ are random. By locality, the choice of $\hat{a}$ does not affect that of $\hat{b}$ and viceversa. Let $A=A(\hat{a},\hat{b}; \psi,\lambda)$ and $B=B(\hat{b},\hat{a};\psi,\lambda)$ be functions which give the results of the measurements on the spins 1 and 2 respectively; for simplicity we normalize their values to $\pm 1$. Locality is taken into account if $A=A(\hat{a}; \psi,\lambda)$ and $B=B(\hat{b};\psi,\lambda)$. Bell also asumed that there is a classical statistical distribution of the variables $\lambda$ in the set $\Lambda$, given by a function $\mu(\lambda)$, with $\int_\Lambda d\lambda \mu(\lambda)=1.$ Then the {\it average value of the product of the projections of the spins (correlation)} is given by $$P(\hat{a}, \hat{b};\psi)=\int_\Lambda d\lambda \mu(\lambda)A(\hat{a};\psi,\lambda)B(\hat{b};\psi,\lambda). \eqno{(3.1)}$$ This is {\it not} a consequence of only QM, but of QM and the hypotesis ii) and iii) (locality and realism) of EPR, where realism is represented by the set of variables $\lambda$. For $\hat{b}=\hat{a}$ one has the situation of EPR: perfect correlation, and $P(\hat{a},\hat{a};\psi)=-1$; so $$0=\int_\Lambda d\lambda \mu(\lambda)(A(\hat{a};\psi,\lambda)B(\hat{a};\psi,\lambda)+1);$$ then (except possibly for a set of measure zero), $A(\hat{a};\psi,\lambda)B(\hat{a};\psi,\lambda)=-1$ {\it i.e.} $B(\hat{a};\psi,\lambda)=-A(\hat{a};\psi,\lambda)$. Then $$P(\hat{a},\hat{b};\psi)=-\int_\Lambda d\lambda\mu(\lambda)A(\hat{a};\psi,\lambda)A(\hat{b};\psi,\lambda).$$ Consider a third direction $\hat{c}$; then $$P(\hat{a},\hat{b};\psi)-P(\hat{a},\hat{c};\psi)=-\int_\Lambda d\lambda\mu(\lambda)(A(\hat{a};\psi,\lambda)A(\hat{b};\psi,\lambda)-A(\hat{a};\psi,\lambda)A(\hat{c};\psi,\lambda))$$ $$=-\int_\Lambda d\lambda \mu(\lambda)A(\hat{a};\psi,\lambda)A(\hat{b};\psi,\lambda)(1-A(\hat{b};\psi,\lambda)A(\hat{c};\psi,\lambda))$$ and so $$\vert P(\hat{a},\hat{b};\psi)-P(\hat{a},\hat{c};\psi)\vert \leq\int_\Lambda d\lambda\mu(\lambda)\vert 1-A(\hat{b};\psi,\lambda)A(\hat{c};\psi,\lambda)\vert =\int_\Lambda d\lambda \mu(\lambda)(1-A(\hat{b};\psi,\lambda)A(\hat{c};\psi,\lambda))$$ $$=1-\int_\Lambda d\lambda \mu(\lambda)A(\hat{b};\psi,\lambda)A(\hat{c};\psi,\lambda)$$ {\it i.e.} $$\vert P(\hat{a},\hat{b};\psi)-P(\hat{a},\hat{c};\psi)\vert \leq 1+P(\hat{b},\hat{c};\psi). \eqno{(3.2)}$$ This is the simplest {\it Bell's inequality} (Bell, 1964). Notice that it does not depend on $\lambda's$ since these variables have been integrated. 

The quantum prediction for $P(\hat{a},\hat{b};\psi)$, which we denote by $P_q(\hat{a},\hat{b};\psi)$, is given by $$P_q(\hat{a},\hat{b};\psi)=-\hat{a}\cdot\hat{b} \eqno{(3.3)}$$ (see Appendix). We prove now that (3.2) and (3.3) are contradictory: (3.2) amounts to $$-1-P(\hat{b},\hat{c};\psi)\leq P(\hat{a},\hat{b};\psi)-P(\hat{a},\hat{c};\psi)\leq 1+P(\hat{b},\hat{c};\psi);$$ taking $\hat{a}$, $\hat{b}$ and $\hat{c}$ in a plane, with $\widehat{(\hat{a},\hat{b})}=\widehat{(\hat{b},\hat{c})}={{1}\over{2}}\widehat{(\hat{a},\hat{c})}=\theta\in[0,{{\pi}\over{2}}]$, one obtains $P_q(\hat{a},\hat{b};\psi)=P_q(\hat{b},\hat{c};\psi)=-cos\theta$, $P_q(\hat{a},\hat{c};\psi)=-cos2\theta$. If QM reproduces Bell-EPR, then one should have $$-1+cos\theta \leq -cos\theta+cos2\theta \leq 1-cos\theta.$$ The second inequality holds since $cos2\theta \leq 1$; however, for the first inequality, $2cos\theta \leq 1+cos2\theta \leq 1+cos^2\theta-sin^2\theta=2cos^2\theta$ {\it i.e.} $cos\theta \leq cos^2\theta$. If $\theta={{\pi}\over{2}}$ (0) then 0$\leq$0 (1$\leq$1), but if $\theta\in (0,{{\pi}\over{2}})$ then $$1\leq cos\theta \eqno{(3.4)}$$ which is {\it false}. Then, {\it the Bell-EPR hypotesis can not reproduce the predictions of QM.}    QED

\ 

Notice that for the case of {\it perfect correlation}, $\hat{a}=\hat{b}$ and then $$P(\hat{a},\hat{a};\psi)=P_q(\hat{a},\hat{a};\psi)=-1 \eqno{(3.5)}$$ {\it i.e.} EPR-Bell reproduces the QM result. 

\

In 1969, Clauser {\it et al}$^{14}$ derived another inequality which proved to be extremely useful for experimental purposes. If $x,y,x^\prime$ and $y^\prime$ take values in $\{+1,-1\}$, then the following equality holds: $$xy-xy^\prime +x^\prime y+x^\prime y^\prime=\pm 2. \eqno{(3.6)}$$ In fact, the left hand side of (3.6) can be written as $$x(y-y^\prime)+x^\prime(y+y^\prime)$$ and $y-y^\prime$ or $y+y^\prime$ equals $\pm 2$. Identifying: $$x=A(\hat{a};\psi,\lambda), \ y=B(\hat{b};\psi,\lambda), \ x^\prime=A(\hat{a^\prime};\psi,\lambda) \ and \ y^\prime=B(\hat{b^\prime};\psi,\lambda),$$ multiplying by $\mu(\lambda)$, and integrating over $\lambda$, one has the CNSH inequality: $$-2\leq P(\hat{a},\hat{b};\psi)-P(\hat{a},\hat{b^\prime};\psi)+P(\hat{a^\prime},\hat{b};\psi)+P(\hat{a^\prime},\hat{b^\prime};\psi)\leq 2 \eqno{(3.7)}$$ (since $\vert\int\mu(xy-xy^\prime+x^\prime y+x^\prime y^\prime)\vert \leq \int \mu\vert xy-xy^\prime +x^\prime y+x^\prime y^\prime \vert=2\int\mu=2)$. 

It is easy to see how (3.7) is violated by QM$^{15}$: choose the four unit vectors $\hat{b^\prime}$, $\hat{a^\prime}$, $\hat{b}$ and $\hat{a}$ in a plane with (say counterclockwise) angles given by $\widehat{(\hat{b^\prime},\hat{a^\prime})}=\widehat{(\hat{a^\prime},\hat{b})}=\widehat{(\hat{b},\hat{a})}={{\pi}\over{4}}$; then, replacing $P's$ by $P_q's$, for the middle term in (3.7) one obtains $$-cos{{\pi}\over{4}}+cos{{3}\over{4}}\pi-cos{{\pi}\over{4}}-cos{{\pi}\over{4}}=-2\sqrt{2},$$ which violates the inequality.

\

Almost all the experimental results confirm the violation by QM of the above Bell's inequalities and all other inequalities which have been obtained afterwards. A summary of the experimental situation can be found in Aspect$^{16}$, Zeilinger$^{17}$, and Weinfurter$^{18}$. 

In (2.2.g.) we have shown that QM is non local ({\it a` la} EPR), in agreement with the theoretical and experimental violation of the Bell's inequalities. The question of its completeness or not is solved in section 4.

\

{\bf 4. Bell's theorem without inequalities: Greenberger-Horne-Zeilinger$^{19,20}$ (1989)}

\

The GHZ's theorem says that {\it the EPR hypotesis contradict quantum mechanics.}

\

We shall follow the presentation of Mermin$^{21,22}$. 

\

{\it Proof of the GHZ's theorem}

\

Consider {\it three} spin ${{1}\over{2}}$ particles, 1, 2 and 3, which are the products of the decay of an initial particle (source). The Hilbert space ${\cal H}$ of the spin part of the decaying particles is isomorphic to $(\C ^2)^3\cong \C^8$. After the decay, the three particles move freely (and therefore without interaction) along straight lines at $120^\circ$ from each other, in a horizontal plane $yz$ (one approximates the orbital motion by classical trajectories). For $k=1,2,3$, we call $z_k$ the direction of motion of particle $k$, $x_1=x_2=x_3=x$ the direction normal to the plane, and $y_k$
the direction in the plane normal to $z_k$. Along each path $z_k$, one sets two SG apparatuses to measure the spin projections in the directions $x$ and $y_k$. Let $\uparrow_k$ ($\downarrow_k$) be the eigenstates of the spin operator $\sigma_{z_k}$ with eigenvalue +1(-1), and let the normalized state vector of the system of the decaying three particles be given by $$\vert\psi>={{1}\over{\sqrt{2}}}(\vert\uparrow_1\uparrow_2\uparrow_3>-\vert\downarrow_1\downarrow_2\downarrow_3>). \eqno{(4.1)}$$ $\vert\psi>$ is symmetric under the interchange of the particles. 

\

Consider the spin operators $$A=\sigma_{x_1}\sigma_{y_2}\sigma_{y_3}, \ B=\sigma_{y_1}\sigma_{x_2}\sigma_{y_3}, \ C=\sigma_{y_1}\sigma_{y_2}\sigma_{x_3}. \eqno{(4.2)}$$ They have the following properties:

i) Hermiticity: $$A^\dagger=(\sigma_{x_1}\sigma_{y_2}\sigma_{y_3})^\dagger=\sigma_{y_3}^\dagger\sigma_{y_2}^\dagger\sigma_{x_1}^\dagger=\sigma_{y_3}\sigma_{y_2}\sigma_{x_1}=\sigma_{x_1}\sigma_{y_2}\sigma_{y_3}=A, \eqno{(4.3)}$$ since spin operators for different particles commute with each other; then also $$B^\dagger=B \ and \ C^\dagger=C. \eqno{(4.3')}$$

ii) Commutativity: $$AB=\sigma_{x_1}\sigma_{y_2}\sigma_{y_3}\sigma_{y_1}\sigma_{x_2}\sigma_{y_3}=(-1)\sigma_{y_1}\sigma_{x_1}\sigma_{y_2}\sigma_{y_3}\sigma_{x_2}\sigma_{y_3}=(-1)^2\sigma_{y_1}\sigma_{x_2}\sigma_{y_3}\sigma_{x_1}\sigma_{y_2}\sigma_{y_3}=BA, \eqno{(4.4)}$$ etc. 

iii) Eigenvalues $\pm 1$: $$A^2=B^2=C^2=1 \eqno{(4.5)}$$ since e.g. $$A^2=\sigma_{x_1}\sigma_{y_2}\sigma_{y_3}\sigma_{x_1}\sigma_{y_2}\sigma_{y_3}=\sigma_{x_1}^2\sigma_{y_2}^2\sigma_{y_3}^2=1.$$ Then the set of eight eigenstates of $A$, $B$ and $C$, $$\{\vert 1,1,1>,\vert -1,1,1>, \vert 1,-1,1>, \vert 1,1,-1>, \vert 1,-1,-1>, \vert -1,1,-1>, \vert -1,-1,1>, \vert -1,-1,-1>\} \eqno{(4.6)}$$ form a basis of ${\cal H}$ and therefore $A$, $B$ and $C$ is a complete set of commuting observables of the particles 1, 2 and 3. (The entries in $\vert a,b,c>$, with $a,b,c=\pm1$, respectively are the eigenvalues of $A$, $B$ and $C$.)

iv) $\vert \psi>$ is eigenstate of $A$, $B$ and $C$ with eigenvalue equal to 1 {\it i.e.} $$A\vert\psi>=B\vert\psi>=C\vert\psi>=\vert \psi>. \eqno{(4.7)}$$ This can be easily verified taking into account that for any of the three particles, $\sigma_{x_k}\vert\uparrow_k>=\vert\downarrow_k>$, $\sigma_{x_k}\vert \downarrow_k>=\vert\uparrow_k>$, $\sigma_{y_k}\vert\uparrow_k>=i\vert\downarrow_k>$, $\sigma_{y_k}\vert\downarrow_k>=-i\vert\uparrow_k>$ (in fact $\pmatrix{0 & 1 \cr 1 & 0 \cr}\pmatrix{1 \cr 0 \cr}=\pmatrix{0 \cr 1 \cr}$, etc.) Then, $$A\vert\psi>={{1}\over{\sqrt{2}}}(\sigma_{x_1}\sigma_{y_2}\sigma_{y_3}\vert \uparrow_1\uparrow_2\uparrow_3>-\sigma_{x_1}\sigma_{y_2}\sigma_{y_3}\vert\downarrow_1\downarrow_2\downarrow_3>)={{1}\over{\sqrt{2}}}(i^2\vert\downarrow_1\downarrow_2\downarrow_3>-(-i)^2\vert\uparrow_1\uparrow_2\uparrow_3>)=\vert\psi>,$$ etc. 

\

Suppose that on particles 3 and 2 one measures the spin projections with SG's in directions $y_3$ and $y_2$ and obtain, say, the values +1 and -1 respectively. Then one can predict with certainty (perfect correlation) that the spin projection of particle 1 in the $x_1$ direction is -1 since the state $\vert \psi>$ is an eigenstate of $A=\sigma_{x_1}\sigma_{y_2}\sigma_{y_3}$ with eigenvalue +1, and so the product of the three spin projections must be +1. Assuming {\it locality} (EPR hypotesis ii) in section 2), the measurements on particles 2 and 3 do not disturb particle 1 since the particles are far enough so that they do not interact with each other. Then according to the criterion iii) of EPR (section 2), the eigenvalue -1 of $\sigma_{x_1}$ is an {\it element of physical reality}. The same analysis can be repeated for other measurements and one concludes with EPR that the six eigenvalues of $\sigma_{x_1}$, $\sigma_{y_1}$, $\sigma_{x_2}$, $\sigma_{y_2}$, $\sigma_{x_3}$, $\sigma_{y_3}$, respectively $m_{x_1}$, $m_{y_1}$, $m_{x_2}$, $m_{y_2}$, $m_{x_3}$, $m_{y_3}$ $\in\{1,-1\}$ are elements of the physical reality. Since $\sigma_{x_k}$ and $\sigma_{y_k}$ for $k=1,2,3$ can not have a common eigenvector,   this leads to the conclusions that, as in the case of EPR, QM is not a complete theory, or, if it is complete, then it is in conflict with local realism. Since locality is violated (2.2.g), there remain the two possibilities: QM is complete or realist. 

Assuming however the existence of the above elements of reality, they obey $$m_{x_1}m_{x_2}m_{x_3}=1 \eqno{(4.8)}$$ since $1=m_{x_1}m_{y_2}m_{y_3}m_{y_1}m_{x_2}m_{y_3}m_{y_1}m_{y_2}m_{x_3}=m_{x_1}m_{x_2}m_{x_3}$. But then consider the operator $$D=\sigma_{x_1}\sigma_{x_2}\sigma_{x_3}. \eqno{(4.9)}$$ D commutes with $A$, $B$ and $C$: in fact $$AD=\sigma_{x_1}\sigma_{y_2}\sigma_{y_3}\sigma_{x_1}\sigma_{x_2}\sigma_{x_3}=\sigma_{x_1}\sigma_{x_1}\sigma_{y_2}\sigma_{y_3}\sigma_{x_2}\sigma_{x_3}=(-1)\sigma_{x_1}\sigma_{x_2}\sigma_{x_1}\sigma_{y_2}\sigma_{y_3}\sigma_{x_3}=(-1)^2\sigma_{x_1}\sigma_{x_2}\sigma_{x_3}\sigma_{x_1}\sigma_{y_2}\sigma_{y_3}$$
$=DA$, etc. Since $A$, $B$ and $C$ are a complete set of observables then $\vert\psi>$ is also an eigenstate of $D$ with eigenvalue, according to EPR, equal to $m_{x_1}m_{x_2}m_{x_3}=1$. However, as it can be easily verified, $$D=-ABC, \eqno{(4.10)}$$ (in fact $-\sigma_{x_1}\sigma_{y_2}\sigma_{y_3}\sigma_{y_1}\sigma_{x_2}\sigma_{y_3}\sigma_{y_1}\sigma_{y_2}\sigma_{x_3}=-(-1)(\sigma_{y_2})^2(\sigma_{y_3})^2(\sigma_{y_1})^2\sigma_{x_1}\sigma_{x_2}\sigma_{x_3}=\sigma_{x_1}\sigma_{x_2}\sigma_{x_3}$) and then according to QM, $$D\vert\psi>=-ABC\vert\psi> \eqno{(4.11)}$$ {\it i.e.} $\vert\psi>$ is an eigenvector of $D$ with eigenvalue -1, contrary to the prediction of EPR {\it i.e.} QM contradicts realism.   QED

Corollary: {\it QM is a complete theory.}

\

There is an argument in favor of the statement: {\it either QM is incomplete or it is non local}, which in particular implies that {\it if QM is complete then it is non local}. The argument is based on the so called ``Einstein boxes''$^{11,23}$: If a box initially containing a quantum particle is separated into two boxes $A$ and $B$, then the wave function of the particle becomes $\vert\psi>={{1}\over{\sqrt{2}}}(\vert A>+\vert B>)$, where $<A\vert \psi>$ ($<B\vert \psi>$) gives the probability amplitude to find the particle in box $A$ ($B$). If at $B$ an observer finds (does not find) the particle, he can not however distinguish between the folowing two alternatives:

i) $A$ did not measure

ii) $A$ measured but did not find (found) the particle. 

This is the analogous situation to that discussed in 2.2.g. Then the observer at $A$ can not send an instantaneous (and therefore superluminal) information (or signal) to the observer at $B$ through this mechanism. A density matrix analysis analogous to that in 2.2.g. can be made here which shows that, even if no instantaneous signal can be transmitted, there is a collapse $\hat{\rho}_B^{red}\to \vert B>$ (or $\hat{\rho}_A^{red}\to \vert A>$. In fact, $$\hat{\rho}=\vert\psi>\otimes <\psi\vert={{1}\over{2}}(\vert A>\otimes<A\vert + \vert B>\otimes<B\vert +\vert A>\otimes <B\vert +\vert B>\otimes <A\vert );$$ $$\hat{\rho}_A^{red}=\hat{\rho}_B^{red}=tr_A\hat{\rho}=tr_B\hat{\rho}={{1}\over{2}}I$$ which imply $$(\hat{\rho}_A^{red})^2=(\hat{\rho}_B^{red})^2={{1}\over{4}}I\neq {{1}\over{2}}I,$$ $$p(A)=p(B)=<A\vert \hat{\rho}_A^{red}\vert A>=<B\vert \hat{\rho}_B^{red}\vert B>={{1}\over{2}}.$$ It is clear that $p(A)$ and $p(B)$ respectively are the joint probabilities $p(A,-B)$ and $p(-A,B)$ where $-B$ ($-A$) means that the particle is not found at $B$ ($A$). Then for the conditional probabilities: $$p(-A\vert B)={{p(-A,B)}\over{p(B)}}={{p(B)}\over {p(B)}}=1,$$ $$p(A\vert -B)={{p(A,-B)}\over{p(A)}}={{p(A)}\over{p(A)}}=1$$ (certainty).       

\

{\bf 5. Hardy's theorem}$^{24}$ {\bf (1992)}

\

The Hardy's theorem has two parts:

a) {\it QM contradicts local realism.}

b) {\it Elements of reality a` la EPR, corresponding to Lorentz invariant observables, are not Lorentz invariant.}

The main interest for the 2nd. part of the theorem, was the possibility of the existence of a prefered reference frame -like the microwave background radiation- to avoid paradoxes like going backwards in time if superluminal signals or instantaneous actions at a distance are allowed. However, as discussed in subsections 2.2.g. and 2.2.h., even if there are non local instantaneous effects, they do not consist in sending signals or information at velocities greater than $c$, and so paradoxes associated to that possibility are abscent.

\

5.1. {\it Mach-Zehnder interferometer}

The scheme of the MZ apparatus is given in Figure 1. We describe it here in terms of photons, but the treatment for electrons and positrons in Hardy's theorem (subsection 5.2.) is similar. 

$BS_k$, $k=1,2$, are beam-splitters (half-silvered mirrors in the case of photons) and $M_l$, $l=1,2$ are totally reflecting mirrors; $G$ and $F$ are detectors. $a$, $b$, $c$, $d$, $e$, $f$ and $g$ are the quantum states (vectors in a Hilbert space) of a particle taking the corresponding "path". (A complete treatment should use a path integral.) As usual, a reflection on a splitter or a mirror introduces a 90º phase (factors i). Consider the following two possibilities: 

i) With $BS_2$ present

The succesive quantum states of the incident particle are given by the following chain: $$a \buildrel \ {BS_1} \over \longrightarrow {{1}\over {\sqrt{2}}}(b+ic) \buildrel \ {M_2,M_1} \over \longrightarrow {{1} \over {\sqrt{2}}}(id+i(ie))\buildrel \ {BS_2} \over \longrightarrow {{1}\over {\sqrt{2}}}(i{{1}\over {\sqrt{2}}}(g+if)-{{1}\over {\sqrt{2}}}(f+ig))=-f \eqno{(5.1)}$$ which means that only the detector $F$ clicks {\it i.e.} one has constructive interference al $F$ and destructive interference at $G$: the particle ``goes through'' two paths ({\it zwei Wegs}); in other words, the presence of $BS_2$ makes the particle to exhibit a wave-like nature. Notice that the detector which clicks is the one in the direction of the incident particle. In the analogy with the two slits Young experiment$^{25}$, the situation is equivalent to that with the two slits opened. 

ii) With $BS_2$ removed

$$a\to \cdot \cdot \cdot {{1}\over{\sqrt{2}}}(id-e)\to {{1}\over{\sqrt{2}}}(ig-f) \eqno{(5.2)}$$ which means that both detectors $G$ and $F$ click, each with probability ${{1}\over{2}}$. If $G$ ($F$) clicks, one knows that the particle ``went through'' the path $b-d-g$ ($c-e-f$) ({\it welches Weg}) exhibiting a particle-like nature. This is analogous to block one of the two slits in the Young double slit experiment (no interference). 

As Wheeler$^{26}$ has realized, $BS_2$ can be inserted or removed at the last instants of the experiment {\it i.e.} much later than the moment in which the particle enters the interferometer; then {\it one can decide whether the particle goes through one path (particle-like behavior) or through two paths (wave-like behavior) after the particle has gone through one of these two alternatives!} But this means that one is acting on the past!? 

\

5.2. {\it Proof of Hardy's theorem}

\

Consider the scheme in Figure 2, where there are two MZ interferometers, one for positrons (+) and one for electrons (-). 

5.2.1. The possibility of electron-positron annihilation at $P$, assumed with probability 1 if the electron and positron travel through the intersecting paths, allows the detection of particles at $G^+$ and $G^-$ (interference destroying alternative) with or without the beam splitters $BS_2^{\pm}$ installed. It is clear that if the detection at $G^+$ and $G^-$ occurs, then annihilation at $P$ has not taken place; this is an example of a counterfactual event. Then, from a semiclassical point of view, if the positron path was $a^+\to c^+\to e^+$, then the electron path was $a^-\to b^-\to d^-$, and if the electron path was $a^- \to c^- \to e^-$, then the positron path was $a^+ \to b^+ \to d^+$.

As in 5.1., we follow the succesive quantum states of the incident particles: $$a^{\pm}\buildrel \ {BS_1^{\pm}} \over \longrightarrow {{1}\over{\sqrt{2}}}(b^{\pm}+ic^{\pm});$$the initial state is $a^+a^-$ ($=\vert a^+,a^->=\vert a^+>\otimes \vert a^->$) and therefore $$a^+a^- \buildrel \ {BS_1^{\pm}} \over \longrightarrow {{1}\over{2}}(b^+ +ic^+)(b^- +ic^-)={{1}\over{2}}(b^+b^-+ib^+c^-+ic^+b^--c^+c^-);$$ since $c^+c^-$ annihilate at $P$ {\it i.e.} $$c^+c^-\buildrel \ {P} \over \longrightarrow \gamma's  \ (photons)$$ one has $$a^+a^-\buildrel \ {BS_1^{\pm},P} \over \longrightarrow {{1}\over{2}}(-\gamma's +b^+b^-++ib^+c^-+ic^+b^-)\buildrel \ {M_1^{\pm},M_2^{\pm}} \over \longrightarrow {{1}\over{2}}(-\gamma's-d^+d^--id^+e^--ie^+d^-)\equiv \psi. \eqno{(5.3)}$$ For the beam splitters $BS_2^{\pm}$ we have four possibilities: 

i) Both $BS_2^{\pm}$ active: $$d^{\pm}\buildrel \ {BS_2^{\pm}} \over \longrightarrow {{1}\over{\sqrt{2}}}(if^{\pm}+g^{\pm}), \ e^{\pm}\to {{1}\over {\sqrt{2}}}(f^{\pm}+ig^{\pm});$$ then $$a^+a^-\to \psi \to {{1}\over{4}}(-2\gamma's+3f^+f^--if^+g^--ig^+f^-+g^+g^-). \eqno{(5.4)}$$ Probabilities check: $({{2}\over{4}})^2+({{3}\over {4}})^2+3\times ({{1}\over {4}})^2=1$. 

ii) $BS_2^-$ active and $BS_2^+$ removed: $$d^+\to g^+, \ e^+\to f^+, \ d^-\buildrel \ {BS_2^-} \over \longrightarrow {{1}\over{\sqrt{2}}}(if^-+g^-), \ e^- \buildrel \ {BS_2^-} \over \longrightarrow {{1}\over {\sqrt{2}}}(f^-+ig^-);$$ then $$a\to \psi \to {{1}\over {2\sqrt{2}}}(-\sqrt{2}\gamma's+f^+f^--i(2g^+f^-+f^+g^-)). \eqno{(5.5)}$$ Probabilities check: $({{1}\over{2}})^2+2\times ({{1}\over {2\sqrt{2}}})^2+({{1}\over {2\sqrt{2}}})^2=1$.

iii) $BS_2^+$ active and $BS_2^-$ removed: $$d^-\to g^-, \ e^-\to f^-, \ d^+\buildrel \ {BS_2^+} \over \longrightarrow {{1}\over {\sqrt{2}}}(if^++g^+), \ e^+\buildrel \ {BS_2^+}\over \longrightarrow {{1}\over{\sqrt{2}}}(ig^++f^+);$$ then $$a^+a^-\to \psi \to {{1}\over{2\sqrt{2}}}(-\sqrt{2}\gamma's +f^+f^--i(2f^+g^-+g^+f^-)). \eqno{(5.6)}$$ Probabilities check: $({{1}\over{2}})^2+2\times ({{1}\over{2\sqrt{2}}})^2+({{1}\over{\sqrt{2}}})^2=1$.

iv) Both $BS_2^{\pm}$ removed: $$d^{\pm}\to g^{\pm}, \ e^{\pm}\to f^{\pm};$$ then $$a^+a^-\to \psi \to {{1}\over{2}}(\gamma's+g^+g^-+i(g^+f^-+f^+g^-)). \eqno{(5.7)}$$ Probabilities check: $4\times ({{1}\over{2}})^2=1$. 

The results (5.3)-(5.7) are a prediction of QM. 

\

5.2.2. One assumes {\it locality} and {\it realism} by introducing the following eight functions: $$F^{\pm}(0,\lambda), G^{\pm}(0,\lambda)=\{\matrix{1, \ if \ a \ positron \ (electron) \ is \ detected \ at \ F^{\pm},G^{\pm} \ with \ BS_2^{\pm} \ installed \cr 0, \ if \ a \ positron \ (electron) \ is \ not \ detected \ at \ F^{\pm},G^{\pm} \ with \ BS_2^{\pm} \ installed \cr}$$ $$F^{\pm}(\infty,\lambda), G^{\pm}(\infty,\lambda)=\{\matrix{1, \ if \ a \ positron \ (electron) \ is \ detected \ at \ F^{\pm},G^{\pm} \ with \ BS_2^{\pm} \ removed \cr 0, \ if \ a \ positron \ (electron) \ is \ not \ detected \ at \ F^{\pm},G^{\pm} \ with \ BS_2^{\pm} \ removed \cr}$$ Realism consists in the assumption that these functions of the hidden variable $\lambda$ exist, while locality consists in the assumption that the functions for positrons (electrons) depend only of $BS_2^+$ ($BS_2^-$). Clearly, the $F$ and $G$ functions are the analogous of the $A$ and $B$ functions of Bell, while the beam splitters $BS_2^{\pm}$ play the r$\hat{o}$le of the SG's represented by $\hat{a}$ and $\hat{b}$, in section 3. 

From case iv), no $f^+f^-$ term appears in (5.6), then $$F^+(\infty,\lambda)F^-(\infty,\lambda)=0 \eqno{(5.8)}$$ in all experiments with both $BS_2^{\pm}$ removed.

From case iii), $$ if \ G^+(0,\lambda)=1 \ then \ F^-(\infty,\lambda)=1 \eqno{(5.9)}$$ since the term $g^+f^-$ appears in (5.5); this happens for $({{1}\over{2\sqrt{2}}})^2={{1}\over{8}}$ of the experiments with $BS_2^+$ in place and $BS_2^-$ removed. 

From case ii), $$if \ G^-(0,\lambda)=1 \ then  \ F^+(\infty,\lambda)=1 \eqno{(5.10)}$$ since the term $f^+g^-$ appears in (5.5); this happens for $({{1}\over{2\sqrt{2}}})^2={{1}\over {8}}$ of the experiments with $BS_2^+$ removed and $BS_2^-$ in place. 

From case i), $$G^+(0,\lambda)G^-(0,\lambda)=1 \eqno{(5.11)}$$ for $({{1}\over{4}})^2={{1}\over{16}}$ of the experiments with both $BS_2^{\pm}$ present. 

\

Since $F^{\pm}(0,\lambda)$, $F^{\pm}(\infty,\lambda)$, $G^{\pm}(0,\lambda)$ and $G^{\pm}(\infty,\lambda)$ are functions of $\lambda$, we can compare them independently of the settings $BS_k^{\pm}$ ($k=1,2$) which determine their values. 

From (5.11), $G^+(0,\lambda)=G^-(0,\lambda)=1$, and from (5.9) and (5.10), $F^+(\infty,\lambda)=F^-(\infty,\lambda)=1$ which implies $F^+(\infty,\lambda)F^-(\infty,\lambda)=1$, which is in contradiction with (5.8). Then, {\it QM contradicts local realism}. This is the Bell theorem without inequalities for two particles.

\

5.2.3. Let us assume with Hardy that {\it if an element of physical reality corresponds to a Lorentz invariant observable , then the numerical value of the element of physical reality is itself Lorentz invariant}.  For example, if $\vert a>$  is an eigenstate of the observable $A$ with eigenvalue $a$, {\it i.e.} $\hat{A}\vert a>=a\vert a>$ and $A$ is Lorentz invariant, then $[{\cal A}]$=$a$ is Lorentz invariant, where ${\cal A}$ is the element of physical reality corresponding to $A$ and $[{\cal A}]$ is its value (we have used the EPR sufficient criterion of reality iii) of section 2). 

Define the operators $$\hat{E}^{\pm}=\vert e^{\pm}>\otimes <e^{\pm}\vert , \ \hat{E}=\hat{E}^+\hat{E}^-.\eqno{(5.12)}$$ $\hat{E}^{\pm}$ are Lorentz invariant since they are the projectors onto the corresponding ``arms'' of the MZ's, and $\hat{E}$ is Lorentz invariant since it is the product of Lorentz invariant operators. Since $(\hat{E}^{\pm})^2=\hat{E}^{\pm}$, and $(\hat{E}^+\hat{E}^-)^2=\hat{E}^+\hat{E}^-$ since $\hat{E}^+$ and $\hat{E}^-$ commute, their eigenvalues are 0 and 1. Clearly, $$\hat{E}^{\pm}\vert e^{\pm}>=\vert e^{\pm}>, \ (\hat{E}+\hat{E}^-)\vert e^+,e^->=\vert e^+,e^-> \eqno{(5.13)}$$ and therefore $$[{\cal E}^+]=[{\cal E}^+ {\cal E}^-]=1. \eqno{(5.14)}$$ Also, if $\vert e^+,e^->_\perp$ is a state vector orthogonal to $\vert e^+,e^->$, then $$(\hat{E}^+ \hat{E}^-)\vert e^+,e^->_\perp=0 \ implies \ [{\cal E}^+{\cal E}^-]=0 \eqno{(5.15)}$$ since ${\cal H}\ni 0=0\vert e^+,e^->_\perp$ where in the right hand side $0\in \C$. 

The other way around, if a system has an element of physical reality ${\cal A}$ corresponding to the observable $A$ with value $a$, then the state vector of the system is an eigenvector of the operator $A$ with eigenvalue $a$. 

In particular, $$if \ [{\cal E}^+][{\cal E}^-]=1 \ then \ [{\cal E}^+]=[{\cal E}^-]=1 \ and \ therefore \ [{\cal E}^+{\cal E}^-]=1 \eqno{(5.16)}$$ since the state is $\vert e^+,e^->$ and $(\hat{E}^+\hat{E}^-)\vert e^+,e^->=\hat{E}^+\vert e^+>\otimes \hat{E}^-\vert e^->= \vert e^+>\otimes \vert e^->= \vert e^+,e^->$.

\

Let $K_+$ ($K_-$) be a reference frame in which the positron (electron) is observed before the electron (positron); if the positron (electron) has already passed through $BS_2^+$ ($BS_2^-$) but the electron (positron) has not yet passed through $BS_2^-$ ($BS_2^+$), then the process in the MZ's is given by $$a^+a^-\to {{1}\over{2\sqrt{2}}}(-\sqrt{2}\gamma's-2if^+d^-+f^+e^--ig^+e^-) \eqno{(5.17)}$$ since in (5.3), $d^+\to {{1}\over{\sqrt{2}}}(if^++g^+)$, $e^+\to{{1}\over{\sqrt{2}}}(ig^++f^+)$ but $d^-\to d^-$ and $ e^-\to e^-$ $$(a^+a^-\to {{1}\over{2\sqrt{2}}}(-\sqrt{2}\gamma's-2id^+f^-+e^+f^--ie^+g^-) \eqno{(5.18)}$$ since in (5.3), $d^-\to {{1}\over{\sqrt{2}}}(if^-+g^-)$, $e^-\to{{1}\over{\sqrt{2}}}(f^-+ig^-)$ but $d^+\to d^+$ and $e^+\to e^+)$. 

Then:

\

In $K_+$, if the positron is detected in $G^+$, then the electron collapses to the state $-ie^-$ and therefore $$[{\cal E}^-]=1. \eqno{(5.19)}$$ In $K_-$, if the electron is detected in $G^-$, then the positron collapses to the state $-ie^+$ and therefore $$[{\cal E}^+]=1. \eqno{(5.20)}$$ In $K_0$, it is easy to verify by an explicit calculation that $$(\hat{E}^+\hat{E}^-)\psi=0 \eqno{(5.21)}$$ with $\psi$ given in (5.3); then $$[{\cal E}^+{\cal E}^-]=0. \eqno{(5.22)}$$ Clearly, this result contradicts (5.16) together with (5.19) and (5.20). The contradiction emerges because one is comparing the values of elements of physical reality in different reference frames, assuming that they were Lorentz invariant.    QED

The contradiction can be \ seen in another way: In $K_+$, $[{\cal E}^-]=1$ means that the electron went through $e^-$, then the positron went through $b^+$ to avoid annihilation at $P$; in $K_-$, $[{\cal E}^+]=1$ means that the positron went through $e^+$ and therefore the electron through $b^-$ again to avoid annihilation at $P$. So, in different frames the trajectories are not the same; moreover, the trajectories predicted in $K_+$ and $K_-$ are contradictory if we stay at $K_0$  with both $G^+$ and $G^-$ detecting the particles.

\

$^{a)}$ Based on a talk given by the author at the Simposio de la Sociedad Cubana de F\'\i sica, La Habana, Cuba, May 19th, 2005.

\

{\bf Acknowledgement}

\

The author thanks the graduate student Brenda Carballo for enlightened discussions. This work was partially supported by the project PAPIIT IN103505, DGAPA-UNAM, M\'exico.

\

{\bf References}

\

1. A. Einstein, B. Podolsky and N. Rosen, ``Can Quantum-Mechanical Description of Physical Reality Be Considered Complete?, Phys. Rev. {\bf 47} (1935) 777-780.

\

2. E. Schroedinger, "Discussion of Probability Relations Between Separated Systems'', Proc. Camb. Phil. Soc. {\bf 31} (1935) 555-563.

\

3. A. Einstein, ``Autobiographical Notes'', in Albert Einstein, Philosopher-Scientist, The Library of Living Philosophers, Vol. VII, ed. by P. A. Schilpp, Northwestern University and Southern Illinois University, 1949, p. 81.

\

4. D. Bohm, ``Quantum Theory'', Dover 1989, chapter 22, p. 611.

\

5. N. Bohr, ``Can quantum-mechanical description of physical reality be considered complete?, Phys. Rev. {\bf 48} (1935) 696-702.

\

6. D. M. Greenberger, M. A. Horne, A. Shimony and A. Zeilinger, ``Bell's theorem without inequalities'', Am. J. Phys. {\bf 58} (1990) 1131-1143, note 10. 

\

7. A. Cabello and G. Garc\'\i a Alcaine, ``La sorprendente incompatibilidad de la idea de realidad einsteniana con la mec\'anica cu\'antica'', Rev. Espa$\tilde{n}$ola de F\'\i sica {\bf 9} (1995) 11-17.

\

8. D. N. Page, ``The Einstein-Podolsky-Rosen physical reality is completely described by quantum mechanics'', Phys. Lett. {\bf 91}A (1982) 57-60.

\

9. G. C. Ghirardi, A. Rimini and T. Weber, ``A General Argument against Superluminal Transmission through the Quantum mechanical Measurement Process'', Lett. Nuov. Cim. {\bf 27} (1980) 293-298.

\

10. G. Garc\'\i a Alcaine and G. Alvarez Galindo, ``Las mediciones cu\'anticas no violan la causalidad relativista'', Rev. Espa$\tilde{n}$ola de F\'\i sica {\bf 1} (1987) 29-35.

\

11. L. Hardy, ``Spooky action at a distance in quantum mechanics'', Contemporary Physics {\bf 39} (1998) 419-429. 

\

12. L. E. Ballentine, ``Quantum Mechanics'', World Scientific 2001, pp. 595-598.

\

13. J. S. Bell, ``On the Einstein-Podolsky-Rosen paradox'', Physics {\bf I} (1964) 195-200.

\

14. J. F. Clauser, M. A. Horne, A. Shimony and R. A. Holt, ``Proposed experiment to test local hidden variable theories'', Phys. Rev. Lett. {\bf 23} (1969) 880-884.

\

15. J. F. Clauser and A. Shimony, ``Bell's theorem: experimental tests and implications'', Rep. Prog. Phys. {\bf 41} (1978) 1881-1926.

\

16. A. Aspect, ``Bell's theorem: the naive view of an experimentalist'', in {\it Quantum [Un]speakables-From Bell to Quantum Information}, ed. by R. A. Bertlmann and A. Zeilnger, Springer (2002).

\

17. A. Zeilinger, ``Experiment and the foundations of quantum physics'', Rev. Mod. Phys. {\bf 71} (1999) S288-S297.

\

18. H. Weinfurter, ``The power of entanglement'', Physics World {\bf 18} (2005) 47-51.

\

19. D. M. Greenberger, M. Horne and A. Zeilinger, ``Going beyond Bell's theorem'', in {\it Bell's Theorem, Quantum Theory, and Conception of the Universe}, ed. by M. Kafatos, Kluwer Academic, Dordrecht (1989) 73-76.

\

20. D. M. Greenberger, M. Horne, A. Shimony and A. Zeilinger, ``Bell's theorem without inequalities'', Am. J. Phys. {\bf 58} (1990) 1131-1143.

\

21. N. D. Mermin, ``What's wrong with these elements of reality?'', Physics Today, (june 1990) 9-11.

\

22. N. D. Mermin, ``Quantum mysteries revisited'', Am. J. Phys. {\bf 58} (1990) 731-734.

\

23. T. Norsen, ``Einstein's boxes'', Am. J. Phys. {\bf 73} (2005) 164-176.

\

24. L. Hardy, ``Quantum Mechanics, Local Realistic Theories, and Lorentz Invariant Realistic Theories'', Phys. Rev. Lett. {\bf 68} (1992) 2981-2984.

\

25. R. P. Feynman, R. B. Leighton and M. Sands, ``The Feynman Lectures on Physics'', Vol. III: Quantum Mechanics, Addison-Wesley, Reading, Massachusetts, 1965, chapter 1, pp. 4-9.

\

26. J. A. Wheeler, ``Law without Law'', in Quantum Theory and Measurement, ed. by J. A. Wheeler and W. H. Zurek, Princeton University Press, Princeton, New Jersey, 1983, pp. 182-213. 

\

27. D. Bohm, ``A suggested interpretation of the quantum theory in terms of ``hidden'' variables, I and II'', Phys. Rev. {\bf 85} (1952) 166-193. 

\

{\bf Appendix}

\

{\it Proof of (3.3)}

\

Let $\vert +>=\pmatrix{1 \cr 0 \cr}$ and $\vert ->=\pmatrix{0 \cr 1 \cr}$ be the eigenstates of $\sigma_3=\pmatrix{1 & 0 \cr 0 & -1 \cr}$ with eigenvalues +1 and -1. In a direction $\hat{n}=(sin\theta cos \varphi,sin \theta sin \varphi,cos \theta)$, the normalized eigenvector of $\vec{\sigma}\cdot \hat{n}$ with eigenvalue $+{{1}\over {2}}$ is $\vert \hat{n}>=cos{{\theta}\over{2}}\vert +>+e^{i\varphi}sin{{\theta}\over{2}}\vert ->$. Prepare the singlet state of two particles $\vert \psi>={{1}\over \sqrt{2}}(\vert 1,+>\otimes \vert 2,->-\vert 1,->\otimes \vert 2,+>)$. The probability of finding particle 1 in the +direction of $\sigma_3$ is ${{1}\over {2}}$, then particle 2 will be found with certainty along $z$ with eigenvalue -1; then the probability of finding it in the direction $+\hat{n}$ is $\vert <2,-1\vert 2,\hat{n}>\vert ^2=sin^2{{\theta}\over{2}}$. Then $P_{++}={{1}\over{2}}\times sin^2{{\theta}\over{2}}$. Similarly $P_{--}=P_{++}$ while $P_{+-}=P{-+}={{1}\over{2}}sin^2{{\pi-\theta}\over{2}}={{1}\over{2}}cos^2{{\theta}\over{2}}$. Then $P_q(\hat{z},\hat{n})=(+1)(P_{++}+P_{--})+(-1)(P_{+-}+P_{-+})=sin^2{{\theta}\over{2}}-cos^2{{\theta}\over{2}}=-cos\theta=-\hat{z}\cdot\hat{n}$. By spherical symmetry, this holds for arbitrary $\hat{a}$ and $\hat{b}$.    QED

\

e-mail:

\noindent socolovs@nuclecu.unam.mx

\end